\documentclass[letterpaper,12pt,preprint]{emulateapj}

\def\solmass {\,$\hbox{M}_\odot$}
\def\solum {\,$\hbox{L}_\odot$}
\def\nthp {N$_2$H$^+$}
\def\nhtd {NH$_2$D}
\def\hcop {HCO$^+$}
\def\cs {C$^{34}$S}
\def\kms {km\,s$^{-1}$}

\shorttitle{How Starless Are Starless Cores?}
\shortauthors{Schnee et al.}

\begin{document}

\title{How Starless Are Starless Cores?} 

\author{Scott Schnee\altaffilmark{1}, James Di
  Francesco\altaffilmark{2,3}, Melissa Enoch\altaffilmark{4}, Rachel
  Friesen\altaffilmark{1}, Doug Johnstone\altaffilmark{2,3}, Sarah
  Sadavoy\altaffilmark{3}}

\email{sschnee@nrao.edu}

\altaffiltext{1}{National Radio Astronomy Observatory, 520 Edgemont
  Road, Charlottesville, VA 22903, USA}
\altaffiltext{2}{National Research Council Canada, Herzberg Institute
  of Astrophysics, 5071 West Saanich Road Victoria, BC V9E 2E7,
  Canada}
\altaffiltext{3}{Department of Physics \& Astronomy, University of
  Victoria, Victoria, BC, V8P 1A1, Canada}
\altaffiltext{4}{Department of Astronomy, University of California,
  Berkeley, CA 94720, USA}

\begin{abstract}

In this paper, we present the results of CARMA continuum and spectral
line observations of the dense core Per-Bolo 45.  Although this core
has previously been classified as starless, we find evidence for an
outflow and conclude that Per-Bolo 45 is actually an embedded,
low-luminosity protostar.  We discuss the impact of newly discovered,
low-luminosity, embedded objects in the Perseus molecular cloud on
starless core and protostar lifetimes.  We estimate that the starless
core lifetime has been overestimated by 4-18\% and the Class 0/I
protostellar lifetime has been underestimated by 5-20\%.  Given the
relatively large systematic uncertainties involved in these
calculations, variations on the order of 10\% do not significantly
change either core lifetimes or the expected protostellar luminosity
function.  Finally, we suggest that high resolution (sub)millimeter
surveys of known cores lacking near-infrared and mid-infrared emission
are necessary to make an accurate census of starless cores.

\end{abstract}


\section{Introduction}

Starless cores are the eventual birthplaces of stars, individually or
possibly in small multiples.  They represent the transition between a
diffuse molecular cloud and the next generation of stars to form
therein.  Dense cores are often identified by their dust continuum
emission, and their status as either protostellar or starless can be
classified by the respective presence or absence of an embedded
infrared object or molecular outflow.  For example, cores in nearby
molecular clouds identified by their 1.1\,mm emission have been
classified by \citet{Enoch08}, and cores identified by their
850\,\micron\ emission have been classified by \citet{Hatchell07a},
\citet{Jorgensen07}, and \citet{Sadavoy10}.  Given high amounts of
extinction from dust, near-infrared observations can sometimes miss
deeply embedded protostars, leading to a misclassification of dense
cores.  Reviews of the evolution of low-mass cores can be found in
\citet{DiFrancesco07} and \citet{Ward-Thompson07}.

The first ``starless'' core observed with the {\it Spitzer Space
  Telescope} \citep{Werner04} by the Cores to Disks legacy program
\citep[c2d;][]{Evans03}, L1014, turned out to host an embedded
infrared point source \citep{Young04}.  Other supposedly starless
cores, such as L1448 and L1521F, were also found to harbour Very Low
Luminosity Objects (VeLLOs) that are either protostars or proto-brown
dwarfs \citep{Kauffmann05, Bourke06}.  VeLLOs are defined to be
objects embedded in dense cores that have luminosities $L_{int} <
0.1$\,\solum\ \citep{Kauffmann05,DiFrancesco07}.  A search for VeLLOs
in the c2d survey identified 50 such objects, and 15-25\% of the
``starless'' cores in the survey turned out to be misclassified
\citep{Dunham08}.  Furthermore, \citet{Dunham08} postulated that there
is likely to be a population of embedded objects even fainter than the
sensitivity limit of the c2d survey, leading to the conclusion that
there is still a significant number of hidden VeLLOs.

Despite the progress made with {\it Spitzer}, deeply embedded
protostars will still be missed by near-infrared large-field surveys.
For example, the well-studied protostar IRAS 16293-2422 has not been
detected at wavelengths shorter than 12\,\micron\ \citep{Jorgensen08}.
Recent observations of dense cores with (sub)millimeter
interferometers and deeper mid-infrared observations towards known
cores have found three new VeLLOs or candidate first hydrostatic cores
in the Perseus molecular cloud.  A first hydrostatic core is a
short-lived stage that occurs after collapse has begun, between the
prestellar core and Class 0 protostar phases, in which there is
hydrostatic balance between the thermal pressure of molecular hydrogen
and gravity.  Since first hydrostatic cores qualify as VeLLOs, in this
paper we will group these two classes of objects together unless there
is observational evidence that a particular VeLLO is a first
hydrostatic core.  \citet{Dunham11} have estimated that there should
be between 1-6 first hydrostatic cores in the Perseus molecular cloud,
based on the expected lifetime of first hydrostatic cores \citep[$5
  \times 10^2$ - $5 \times 10^4$
  years;][]{Boss95,Omukai07,Saigo08,Tomida10}, an assumed lifetime of
the embedded protostellar phase \citep[$5.4 \times 10^5$
  years;][]{Evans09}, and the number of embedded prototstars in
Perseus \citep[66;][]{Enoch09}.  Three candidate first hydrostatic
cores have been identified so far in the Perseus molecular cloud.  The
first, L1448 IRS2E, previously believed to be starless based on
non-detections with {\it Spitzer} from 3.6\,$\mu$m to 70\,$\mu$m, was
recently found to host a protostar or first hydrostatic core
\citep{Chen10a}, based on the detection of a collimated CO (2-1)
outflow.  The second, Per-Bolo 58 \citep{Enoch06}, previously believed
to be starless based on {\it Spitzer} observations from the c2d survey
\citep{Evans03}, was found to host an embedded protostar or first
hydrostatic core based on a deep 70\,\micron\ map and 3\,mm continuum
CARMA observations \citep{Enoch10, Schnee10}.  The presence of an
embedded source in Per-Bolo 58 was then confirmed with SMA
observations showing a 1.3\,mm continuum source and CO (2-1) outflow
\citep{Dunham11}.  Third, L1451-mm was also considered starless based
on non-detections in the {\it Spitzer} near-infrared and mid-infrared
images taken as part of the c2d survey.  Follow-up observations of
L1451-mm with the SMA, CARMA, and VLA found compact 3\,mm and 1\,mm
continuum emission coinciding with broadened spectral lines (\nthp,
\nhtd, and NH$_3$) and a slow, poorly collimated CO (2-1) outflow
\citep{Pineda11}.  The discovery that L1448 IRS2E, Per-Bolo 58, and
L1451-mm are not starless suggests that other ``starless'' cores have
been misidentified and instead harbour deeply embedded VeLLOs that
were not detected in wide-area near-infrared and mid-infrared surveys.

In this paper, we present evidence that the dense core Per-Bolo 45
\citep{Enoch06}, despite non-detections in the 3.6-70\,$\mu$m range,
is also protostellar.  Per-Bolo 45 is located in the Perseus molecular
cloud, near the center of the cluster NGC 1333 and between IRAS 7
\citep{Jennings87} to the northeast and SVS 13 \citep{Strom76} to the
southwest.  In \S \ref{OBSERVATIONS} we describe new CARMA spectral
line observations of Per-Bolo 45.  The evidence that Per-Bolo 45 is
protostellar is presented in \S \ref{ANALYSIS}. Some implications of
the discoveries of embedded sources within Per-Bolo 45, Per-Bolo 58,
L1448 IRS2E, and L1451-mm are discussed in \S \ref{DISCUSSION}, and
our results are summarized in \S \ref{SUMMARY}.

\section{Observations} \label{OBSERVATIONS}

Here we present spectral line maps of the dense core Per-Bolo 45 made
with the Combined Array for Research in Millimeter-wave Astronomy
(CARMA).  We made new observations of the 3\,mm transitions of HCN,
\nhtd, SiO, \hcop, HNC, \nthp, and \cs\ at $\sim$5\arcsec\ resolution,
as described in \S \ref{NEWCARMA}.  We discuss these observations in
the context of the 3\,mm continuum data previously published by
\citet{Schnee10} and described briefly in \S \ref{OLDCARMA}.

\subsection{New CARMA Observations}  \label{NEWCARMA}

Spectral line observations in the 3\,mm window were obtained in July
2010 with CARMA, a 15 element interferometer consisting of nine 6.1
meter antennas and six 10.4 meter antennas.  The CARMA correlator
records signals in eight separate bands, each with an upper and lower
sideband.  We configured one band for maximum bandwidth (495\,MHz with
95 channels per band) to observe continuum emission, providing a total
continuum bandwidth of approximately 1\,GHz.  The other seven bands
were configured to observe spectral lines and have 383 channels
covering 7.8\,MHz per band, providing native spectral resolution of
20.4\,kHz (0.067\,\kms), which we smoothed to 0.1\,\kms\ before
analysis.  The observations were centered around 90.7\,GHz, and range
from 85.9\,GHz to 97.9\,GHz.  The half-power beam width of the 10.4\,m
antennas is 74\arcsec\ at the observed frequencies.  Seven-point
mosaics were made toward the dense core Per-Bolo 45 (with the phase
center located at J2000 3:29:07.7 +31:17:16.8, and rest velocity
8.483\,\kms) in the D and E-array configurations, with baselines that
range from 11\,m to 150\,m.  The largest angular scale to which the
observations are sensitive is approximately 30\arcsec.  A summary of
the CARMA observations is presented in Table \ref{MMOBSTAB}.

The observing sequence for the CARMA observations was to integrate on
a phase calibrator (0336+323) for 3 minutes and Per-Bolo 45 for 21
minutes.  The total time spent on this project was about 8 hours in
each of the array configurations.  In each set of observations, 3C84
was observed for passband calibration, and observations of Uranus were
used for absolute flux calibration.  Based on the repeatability of the
quasar fluxes, the estimated random uncertainty in the measured source
fluxes is $\sigma\sim5$\%, and the systematic uncertainty is
approximately 20\%.  Radio pointing was done at the beginning of each
track and pointing constants were updated at least every two hours
thereafter, using either radio or optical pointing routines
\citep{Corder10}.  Calibration and imaging were done using the MIRIAD
data reduction package \citep{Sault95}.

Due to instrumental problems, the first correlator band was lost, so
we have no data on the HCN (1-0) line.  We detected no significant
emission of the C$^{34}$S (2-1) line, and it will not be discussed
further in this paper.

\subsection{Previous Observations} \label{OLDCARMA}

Per-Bolo 45 was included in a CARMA 3\,mm continuum survey of 11
starless cores in the Perseus molecular cloud \citep{Schnee10}.  It is
one of two cores that were detected, and has an integrated 3\,mm flux
density of 11$\pm$0.5\,mJy, corresponding to a mass of
0.8\,\solmass\ after making reasonable assumptions for the core and
dust properties.  See \citet{Schnee10} for more details about the
3\,mm continuum observations and the derived core properties.

\begin{figure}
\epsscale{1.0} 
\plotone{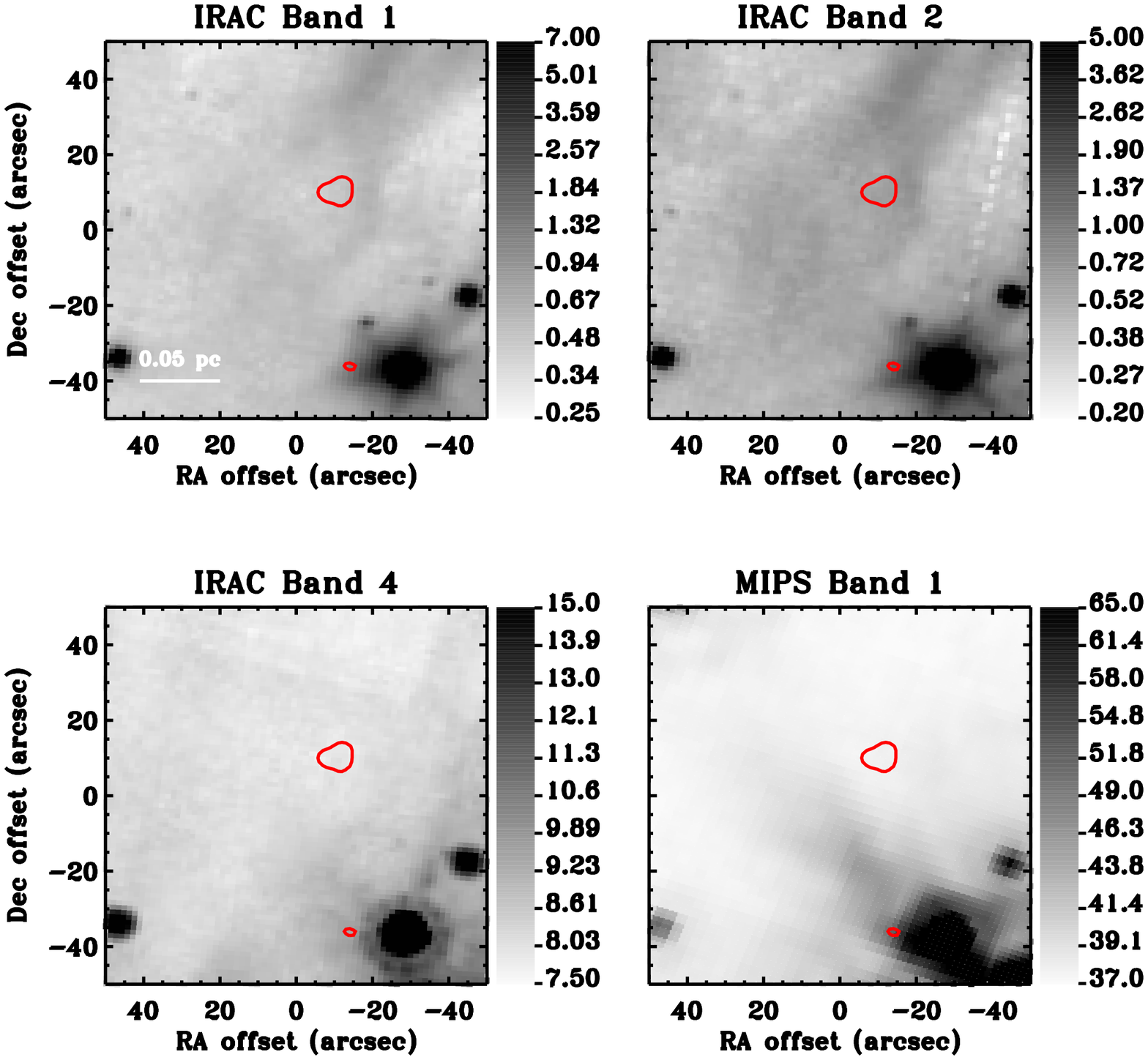}
\caption{{\it Spitzer} near-infrared and mid-infrared maps of Per-Bolo
  45 (greyscale) with CARMA 3\,mm continuum contours.  The {\it
    Spitzer} maps at 3.6\,$\mu$m (top left), 4.5\,$\mu$m (top right),
  8.0\,$\mu$m (bottom left) and 24\,$\mu$m (bottom right) were taken
  as part of the c2d survey \citep{Evans03}.  The 3\,mm continuum data
  come from \citet{Schnee10}, and only the 5\,$\sigma$ flux density
  contour is plotted. The (0,0) position is J2000 3:29:07.7
  +31:17:16.8, and is the same in all figures in this paper.  The
  units of the maps are MJy/sr.
\label{SPITZERMAPS}}
\end{figure}

\begin{figure}
\epsscale{1.0} 
\plotone{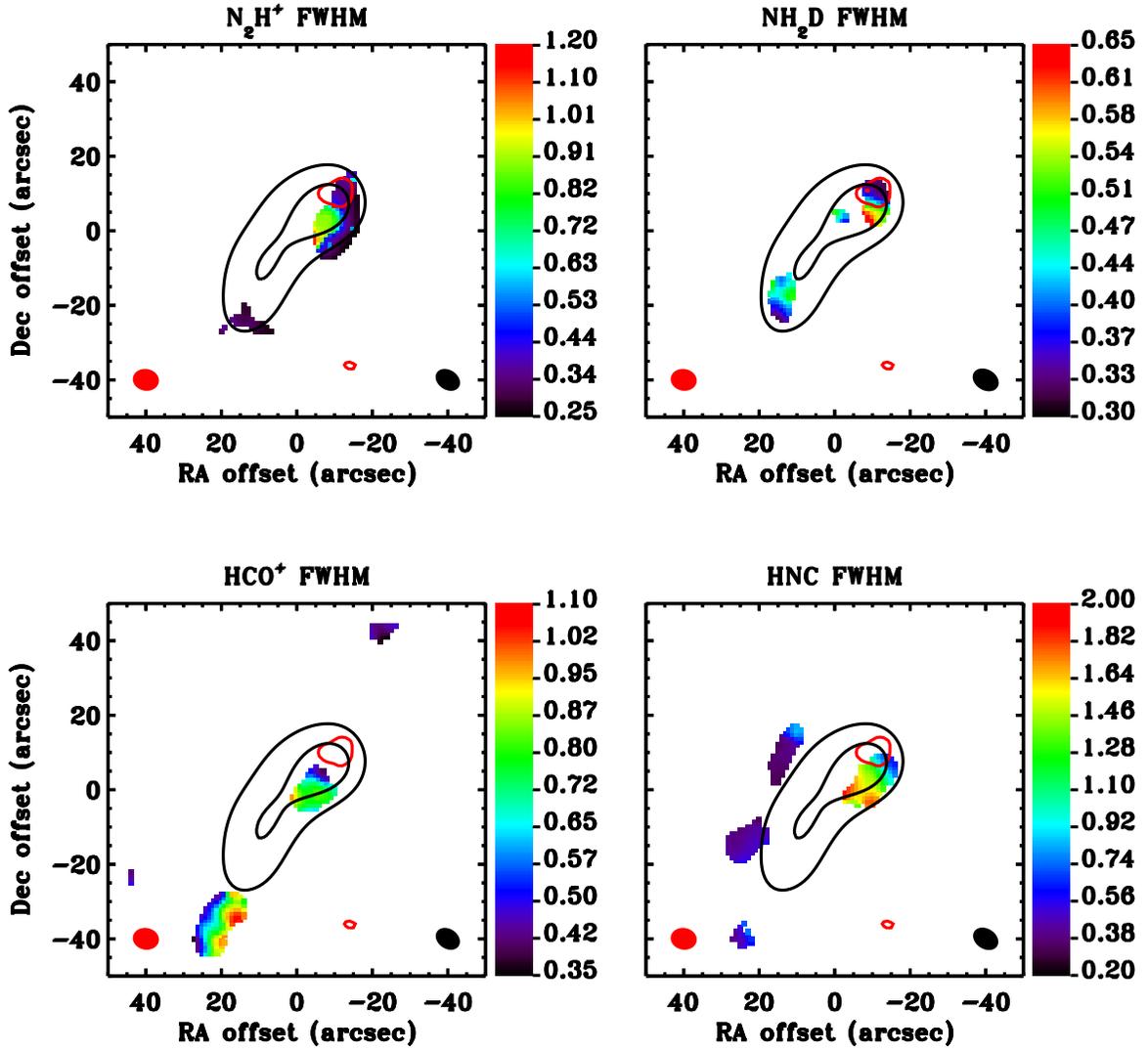}
\caption{The full-width half-maximum (FWHM) of \nthp\ (top left),
  \nhtd\ (top right), \hcop\ (bottom left), and HNC (bottom right).
  The line width is given in color, as shown in the scale bar to the
  right of each panel, in units of \kms.  See \S \ref{ANALYSIS} for an
  explanation of the fits to the spectral line profiles.  Line widths
  are only calculated for positions with at least three channels with
  emission above the 3\,$\sigma$ cutoff.  The red contour shows the
  5\,$\sigma$ dust continuum emission, as in Fig.~\ref{SPITZERMAPS}.
  The black contours show the SCUBA 850\,\micron\ map of Per-Bolo 45
  at 70\% and 90\% of the peak emission in the map
  \citep{DiFrancesco08}. The synthesized beam size of the continuum
  emission is shown in the bottom left corner of each panel, and the
  beam size of the spectral lines is shown in the bottom right corner
  of each panel.
\label{FWHMMAPS}}
\end{figure}

\begin{figure}
\epsscale{1.0} 
\plotone{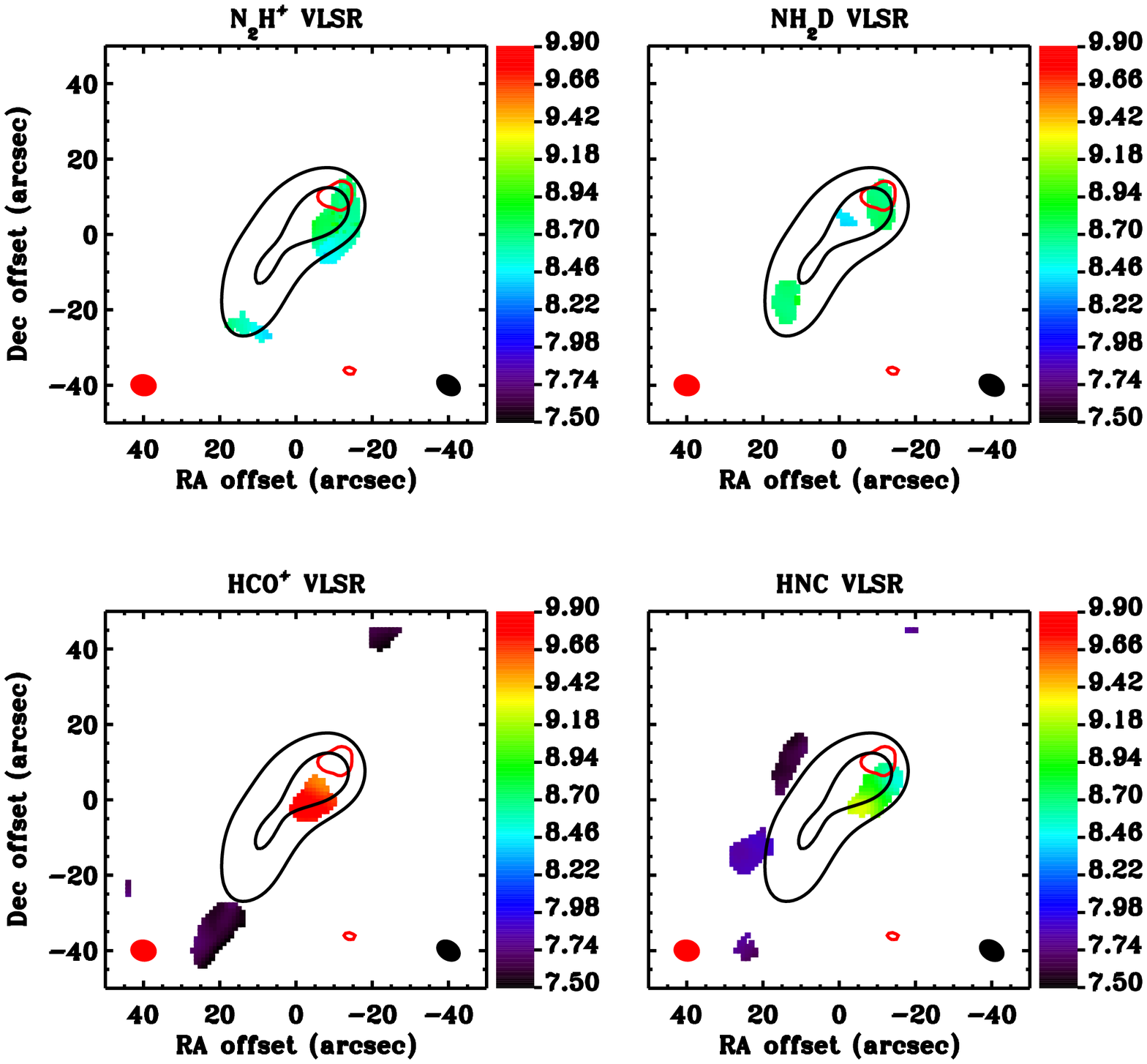}
\caption{The VLSR of \nthp\ (top left), \nhtd\ (top right),
  \hcop\ (bottom left), and HNC (bottom right).  The fit velocity is
  given in color, as shown in the scale bar to the right of each
  panel, in units of \kms.  See \S \ref{ANALYSIS} for an explanation
  of the fits to the spectral line profiles.  Velocities are only
  calculated for positions with at least three channels with emission
  above the 3\,$\sigma$ cutoff.  The red contour shows the 5\,$\sigma$
  dust continuum emission, as in Fig.~\ref{SPITZERMAPS}.  The black
  contours show the SCUBA 850\,\micron\ map of Per-Bolo 45 at 70\% and
  90\% of the peak emission in the map \citep{DiFrancesco08}. The
  synthesized beam size of the continuum emission is shown in the
  bottom left corner of each panel, and the beam size of the spectral
  lines is shown in the bottom right corner of each panel.
\label{VLSRMAPS}}
\end{figure}

\begin{figure}
\epsscale{1.0} 
\plotone{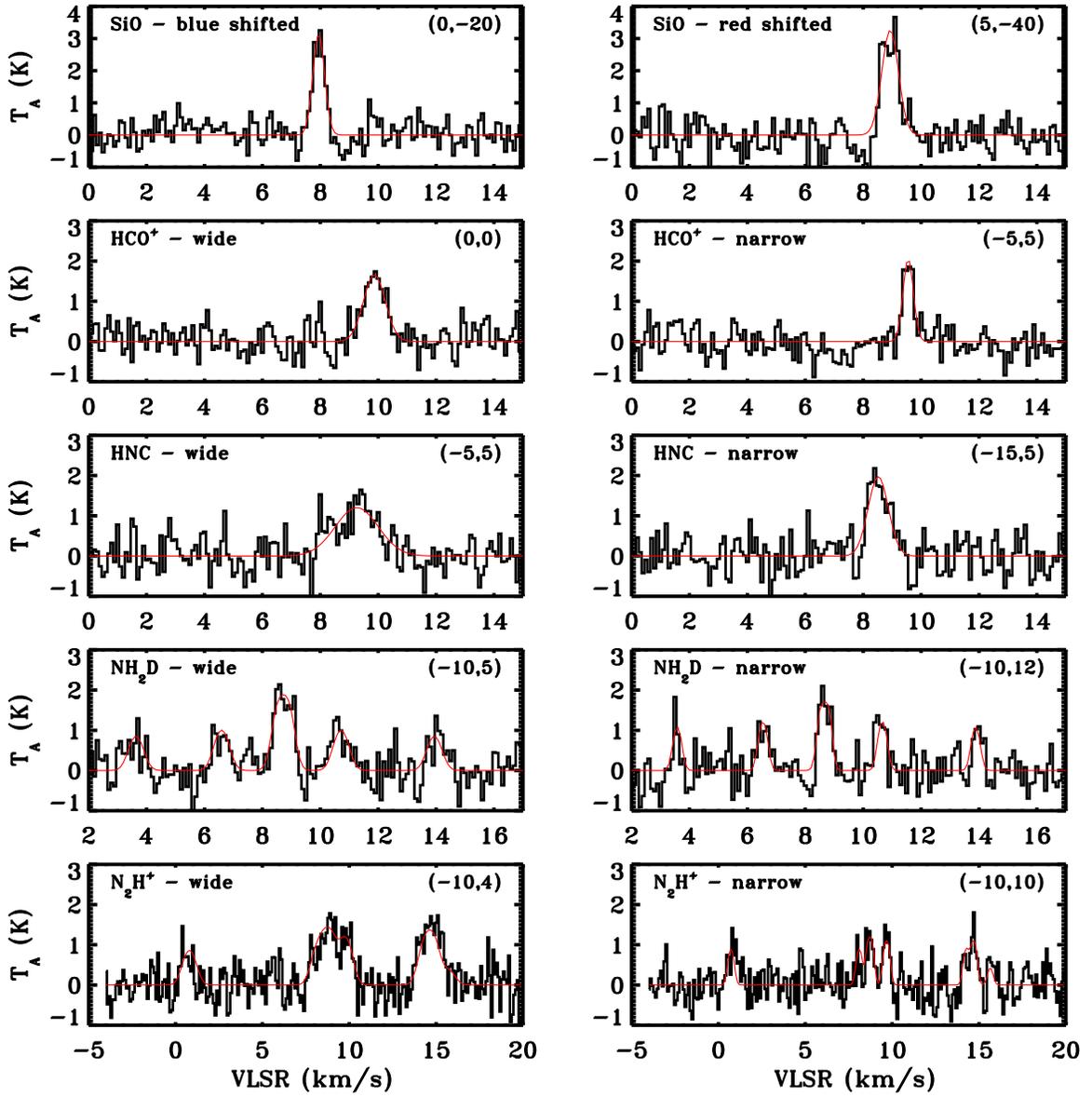}
\caption{Example spectra towards Per-Bolo 45.  SiO emission (top row)
  is presented from a relatively blueshifted position (left) and
  redshifted position (right). For \hcop, HNC, \nhtd, and
  \nthp\ (second from the top to the bottom, respectively), relatively
  wide (left) and narrow (right) line profiles are shown.  The
  positions of the spectra are shown in the top right corner of each
  panel.  Data are shown with the solid black line, and the fit
  profiles are shown with the dotted red line.
\label{SPECTRA}}
\end{figure}

Per-Bolo 45 was classified as a starless core by \citet{Enoch06,
  Enoch08}.  If a protostar was found within one full-width
half-maximum (FWHM) of the center of the 1.1\,mm Bolocam core (i.e.,
31\arcsec, or 0.04\,pc), then it was classified as protostellar,
otherwise the Bolocam core was classified as starless.  Protostars in
\citet{Enoch08} were identified by the shape of their near-infrared to
far-infrared spectral energy distributions (SEDs), a minimum flux at
24\,\micron\ and the presence of 70\,\micron\ point sources not
classified as galaxy candidates.  Per-Bolo 45 was identified as a
Class 0 protostar by \citet{Hatchell07a} based on the colors of a
nearby, but (we believe) unrelated, near-infrared source.  There is a
possible outflow from Per-Bolo 45 reported by \citet{Hatchell07b}, but
due to confusion from nearby protostars this identification is given
as tentative.  In \citet{Jorgensen07}, cores were identified as
protostellar based on either the presence of {\it Spitzer} detections
and colors within a SCUBA core, or by the concentration of the light
profile of the 850\,\micron\ SCUBA flux.  Neither \citet{Jorgensen07},
\citet{Kirk07}, nor \citet{Dunham08} found evidence for an embedded
source in Per-Bolo 45, and it is classified as a starless core by
\citet{Sadavoy10}.  In Fig.~\ref{SPITZERMAPS}, we show {\it Spitzer}
near-infrared and mid-infrared images of Per-Bolo 45 with CARMA 3~mm
continuum overlays, and it is clear that there is no embedded point
source detected by {\it Spitzer} associated within Per-Bolo 45.  The
area shown in Figures \ref{SPITZERMAPS} - \ref{TINTMAPS} is
approximately the region in which the gain of the mosaic is greater
than 0.5.  Therefore, the true classification of Per-Bolo 45 is not
clear from the literature, but most studies have labeled it
``starless.''

\section{Analysis} \label{ANALYSIS}

We study the kinematics and morphology of the dense core Per-Bolo 45
by fitting the line profiles of \nhtd\ (1$_{1,1}$ - 1$_{0,1}$), SiO
(2-1), \hcop\ (1-0), HNC (1-0), and \nthp\ (1-0) using the MPFIT
package in IDL \citep{Markwardt09}.  Single Gaussians were fit to the
SiO, \hcop, and HNC spectra, and multiple Gaussians were fit to the 7
components of the \nthp\ spectra and 6 components of the
\nhtd\ spectra.  Fits were only made to those profiles that had three
independent velocity channels with a signal-to-noise ratio greater
than 3.

\subsection{Molecular Line Emission} \label{MOLECULES}

Starless cores are kinematically quiescent objects, with line widths
broadened approximately equally by thermal and non-thermal motions.
The thermal line widths (FWHM) for the molecular lines presented in
this paper are $\sim$0.13\,\kms, assuming a typical temperature of
10\,K \citep{Schnee09}.  The average non-thermal line width for NH$_3$
in starless cores in the Perseus molecular cloud is greater than the
thermal line width of NH$_3$ by a factor of 1.5 \citep{Foster09}.
Broad line widths (FWHM $\sim$ 0.6\,\kms) in \nthp\ and \nhtd\ towards
the 3\,mm continuum peak of L1451-mm led \citet{Pineda11} to
hypothesize the presence of an embedded protostar in a core that had
previously been considered starless, which was confirmed by detection
of a CO (2-1) outflow.  Similarly, velocity gradients in starless
cores are small, with a typical value of 1 km\,s$^{-1}$\,pc$^{-1}$
\citep{Goodman93}.  We would expect a total velocity gradient across
Per-Bolo 45, which has a diameter no greater than 0.2\,pc, to be
$\sim$0.2\,\kms.  Furthermore, we would expect the velocity field to
be continuous, as seen in interferometric maps of the starless cores
L183 \citep{Kirk09}, L694-2 \citep{Williams06}, and L1544
\citep{Williams06}.  The kinematics of Per-Bolo 45, as traced by
\nthp, \nhtd, \hcop, and HNC, are described in Section \ref{NITROGEN}
and \ref{CARBON}.  Maps of the full-width half-maximum (FWHM) of
\nthp, \nhtd, \hcop, and HNC are shown in Figure \ref{FWHMMAPS}, and
the LSR velocity maps for \nthp, \nhtd, \hcop, and HNC are shown in
Figure \ref{VLSRMAPS}.  Example spectra demonstrating the variety of
line profiles in Per-Bolo 45 are shown in Figure \ref{SPECTRA}.

In the dense, cold, and quiescent interstellar medium, SiO is frozen
onto the surfaces of dust grains except in regions shocked by outflows
that return SiO to the gas phase and excite its emission
\citep{Martin-Pintado92}.  The presence of SiO emission is therefore a
good indicator of one or more nearby protostars.  SiO emission can be
grouped into two classes, one that has broad lines ($>$10\,\kms) and
is Doppler-shifted by velocities $\ge$10\,\kms\ from the ambient
cloud, and a second component with narrow lines ($\sim$1\,\kms) found
at velocities very close to the cloud material
\citep[e.g.,][]{Lefloch98, Jimenez-Serra04}.  The SiO maps of Per-Bolo
45 are discussed in Section \ref{SIO} and presented in Figure
\ref{SIOMAP}.

\begin{figure}
\epsscale{1.0} 
\plotone{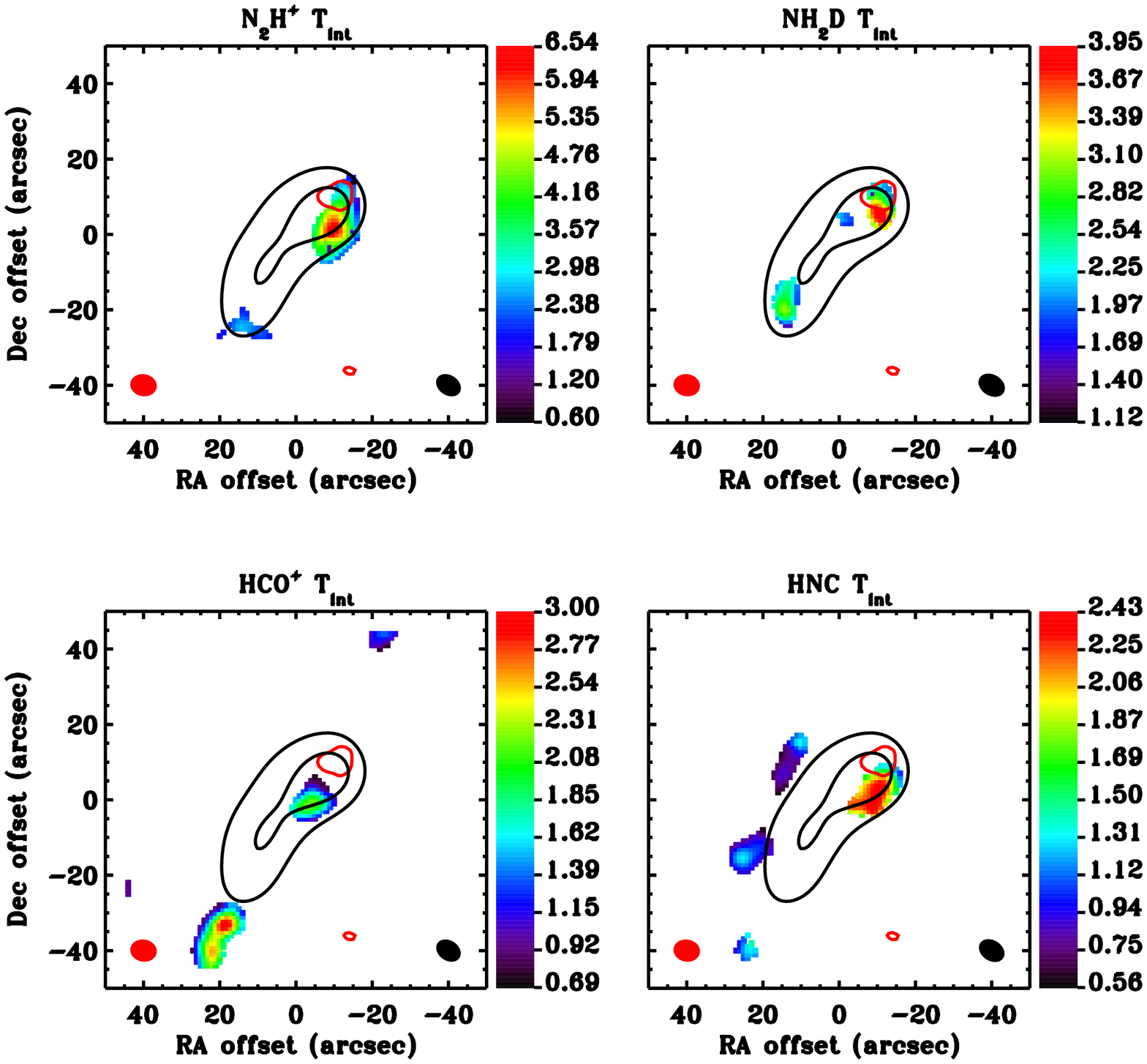}
\caption{The integrated intensity of \nthp\ (top left), \nhtd\ (top
  right), \hcop\ (bottom left), and HNC (bottom right).  The
  integrated intensity is given in color, as shown in the scale bar to
  the right of each panel, in units of K\,\kms.  See \S \ref{ANALYSIS}
  for an explanation of the fits to the spectral line profiles.  The
  red contour shows the 5\,$\sigma$ 3\,mm dust continuum emission, as
  in Fig.~\ref{SPITZERMAPS}.  The black contours show the SCUBA
  850\,\micron\ map of Per-Bolo 45 at 70\% and 90\% of the peak
  emission in the map \citep{DiFrancesco08}.  The synthesized beam
  size of the 3\,mm continuum emission is shown in the bottom left
  corner of each panel, and the beam size of the spectral lines is
  shown in the bottom right corner of each panel.
\label{TINTMAPS}}
\end{figure}

An 850\,\micron\ SCUBA map of the dense core Per-Bolo 45 was produced
as part of SCUBA Legacy Catalogues \citep{DiFrancesco08}, and is shown
in Figure \ref{TINTMAPS} along with the 3\,mm continuum emission and
integrated intensity maps of \nthp, \nhtd, \hcop, and HNC.  Whereas
the 3\,mm continuum emission detected with CARMA has only one peak,
the integrated intensity maps of \nthp, \nhtd, \hcop, and HNC all have
multiple peaks.  The morphology of emission towards Per-Bolo 45 varies
significantly between molecular lines, and this will be discussed
below and presented in Figure \ref{TINTMAPS}.

\subsubsection{\nthp\ and \nhtd} \label{NITROGEN}

The line width of \nthp\ in Per-Bolo 45 has a bimodal distribution,
with narrow lines ($\sim$0.3\,\kms) at the position of the dust peak
and a small region of much broader (0.5-1.2\,\kms) line width a few
arcseconds to the southeast of the dust peak.  The line width of
\nhtd\ is qualitatively similar to that of \nthp, with a narrow line
width at the dust peak and much broader width a few arcseconds to the
southeast.  We argue that the broad line widths (up to 1.2\,\kms) seen
in \nthp\ and \nhtd\ towards Per-Bolo 45 point towards the presence of
a protostar, which must be low luminosity to have not been detected in
mid-infrared {\it Spitzer} maps (see \S\,\ref{EMBEDDED}).  The
velocity fields of \nthp\ and \nhtd\ (as well as \hcop\ and HNC) are
shown in Fig.\,\ref{VLSRMAPS}.  The range of velocities seen in
\nthp\ and \nhtd\ is fairly narrow and similar to the systemic
velocity of the core as measured in the single dish maps.

Both \nthp\ and \nhtd\ are coincident with the 850\,\micron\ dust
emission, but extend about 30\arcsec\ to the southeast beyond the
extent of the 3\,mm continuum.  In both lines the secondary peaks have
narrow line widths typical of quiescent cores, so we suggest that this
material is part of the envelope of Per-Bolo 45 that is not strongly
affected by any outflow.  The secondary peaks of \nthp\ and \nhtd\ are
coincident with the elongation of the 850\,\micron\ dust emission seen
in the SCUBA map, although they are not accompanied by a similar
feature in the 3\,mm continuum map.  Discrepancies on small spatial
scales between the dust emission and dense gas tracers have been seen
before, for instance in the dense core Oph B \citep{Friesen09,
  Friesen10} and in Oph A-N6 \citep{Pon09}.  Furthermore, we show in
this paper that the CARMA continuum maps are not sensitive to the
emission from starless material.

\subsubsection{\hcop\ and HNC} \label{CARBON}

The \hcop\ integrated intensity map (see Fig.~\ref{TINTMAPS}) shows
two peaks, one immediately to the southeast of the dust peak and the
other 40\arcsec\ further to the southeast.  The integrated intensity
map of HNC shows a peak immediately to the southeast of the dust peak
and also has three more peaks to the east and southeast of the 3\,mm
continuum emission.  Two of the HNC emission peaks to the east and
southeast have no counterparts in the dust emission or with the other
features in molecular line maps.  Both \hcop\ and HNC show a broad
range of velocities and both have line widths many times greater than
the thermal line width, as shown in Figs.\,\ref{FWHMMAPS} and
\ref{VLSRMAPS}.  In the case of \hcop, both redshifted and blueshifted
emission are seen with a spread of $\pm$1\,\kms\ from the systemic
velocity of the core, and no emission is seen at the systemic velocity
of Per-Bolo 45.  The HNC emission adjacent to the 3\,mm continuum peak
is at the systemic velocity of the core and has broad line widths (up
to 2\,\kms), while the other HNC peaks are blue-shifted and have
narrow line widths.

Although the interpretation of the emission is complicated by
dissimilarities between the morphologies of the emission, we argue
that the broad line widths (up to 2\,\kms) and velocities
Doppler-shifted from the systemic velocity of the core (by
$\pm$1\,\kms\ in the case of \hcop) are likely caused by an outflow
driven by Per-Bolo 45.  Neither kinematic feature would be expected in
a starless core, as explained in Section \ref{MOLECULES}.  The regions
of narrow line width seen in HNC and \hcop (as well as \nthp\ and
\nhtd) likely trace the more quiescent gas in a dense envelope.
\hcop\ has been seen to trace outflows in the cluster NGC 1333
\citep[e.g.,][]{Walsh07}, of which Per-Bolo 45 is a member.  Given
that HCN has been also seen to trace outflows in NGC 1333
\citep{Jorgensen04}, it is reasonable to suggest that the broad line
width HNC in Per-Bolo 45 traces an outflow as well.  HNC has been seen
to trace dense material around a protostar as well as material
associated with the protostellar outflow \citep{Arce04}, so the
appearance of both quiescent and turbulent HNC in Per-Bolo 45 is not
unexpected if it harbors a protostar.

\subsubsection{SiO} \label{SIO}

\begin{figure}
\epsscale{1.0} 
\plotone{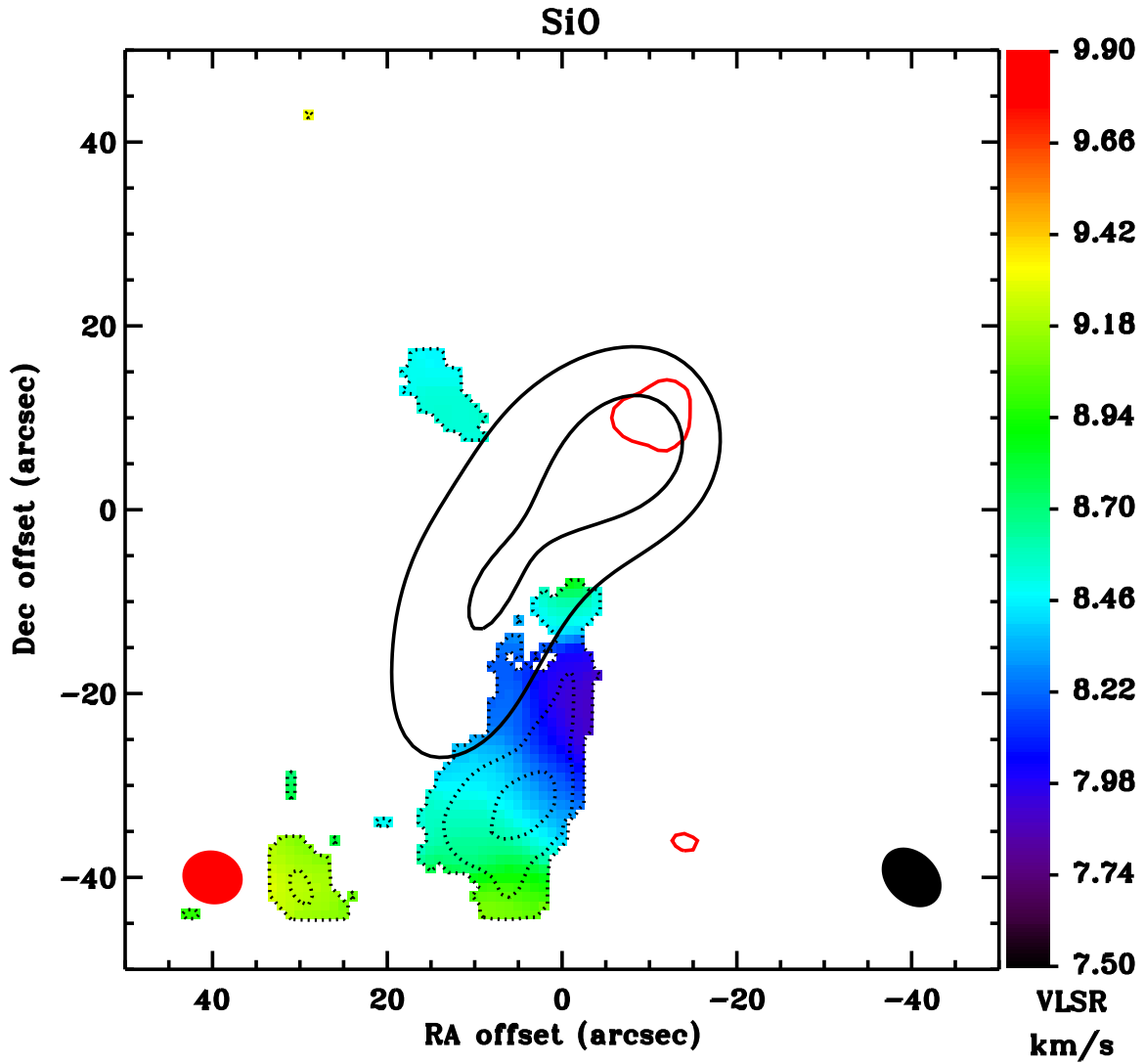}
\caption{SiO (2-1) emission from Per-Bolo 45.  Color indicates the
  VLSR of the emission and dotted black contours show the peak antenna
  temperature (1, 3, 5\,K). The red contour shows the 5\,$\sigma$ dust
  mv continuum emission.  The solid black contours show the SCUBA
  850\,\micron\ map of Per-Bolo 45 at 70\% and 90\% of the peak
  emission in the map \citep{DiFrancesco08}.  The synthesized beam
  size of the continuum emission is shown in red in the bottom left
  corner, and the SiO beam size is shown in black in the bottom right
  corner.  The VLSR of Per-Bolo 45 is 8.5\,\kms\ \citep{Kirk07,
    Rosolowsky08}.  The SiO emission is consistent with gas liberated
  from dust grains and excited by an outflow launched by Per-Bolo 45.
\label{SIOMAP}}
\end{figure}

Per-Bolo 45 exhibits SiO emission, with $\sim$1\,\kms\ line widths
(see Fig.\,\ref{SPECTRA}) and velocities within $\sim$1\,\kms\ of the
ambient medium (see Fig.\,\ref{SIOMAP}).  The systemic velocity of
Per-Bolo 45 is $\sim$8.5\,\kms\ \citep{Kirk07, Rosolowsky08}.  The
systemic velocity of the ISM around Per-Bolo 45, measured in an FCRAO
map of CS (2-1) by the COMPLETE survey \citep{Ridge05}, is 8.6\,\kms,
with a 1.7\,\kms\ line width.  The velocity range of the SiO (2-1)
emission towards Per-Bolo 45 is therefore entirely consistent with the
range of velocities of the ambient material.  The morphology of the
SiO emission is perhaps suggestive, being extended with the long axis
of the emission pointing towards the 3\,mm continuum emission.

We interpret the detection of SiO (2-1) emission immediately adjacent
to Per-Bolo 45 as further evidence for the presence of an embedded
protostar.  An outflow launched by a protostar in Per-Bolo 45 would
liberate SiO from the dust in the ISM, creating the conditions
required for the observed emission.  Since we only detect the narrow
component of SiO emission that traces the ambient cloud material, the
velocity map shown in Fig.\,\ref{SIOMAP} is not sufficient to say for
certain what portion of the outflow is exciting the emission.  The
velocity structure seen in Fig.\,\ref{SIOMAP} is likely that of the
ISM around Per-Bolo 45, and not that of the outflow.  The velocity
range covered by the CARMA observations ($-5$\,\kms\ $\le$ VLSR $\le$
21\,\kms) was not wide enough to find the high-velocity, broad
component of the SiO emission more closely associated with the
outflow, if present.


\subsubsection{The case for the protostellar nature of Per-Bolo 45}

We have shown that Per-Bolo 45 exhibits several traits of protostellar
cores, despite the non-detection of mid-infrared emission in IRAC and
MIPS {\it Spitzer} maps (see Fig.\,\ref{SPITZERMAPS}).  The line
widths seen in \nthp, \nhtd, \hcop, and HNC are several times larger
than the thermal line width, behaviour not found in starless cores but
which could be explained by the presence of an embedded protostar.
Broad line widths of \nthp\ and \nhtd\ in L1451-mm led to the
discovery of a protostar in that core even though it had previously
been identified as starless \citep{Pineda11}.  The presence of
redshifted and blueshifted \hcop\ emission seen about 1\,\kms\ from
the systemic velocity of the core would also not be expected in a
starless core, but could be explained by an outflow from an embedded
protostar.  SiO emission located adjacent to Per-Bolo 45 can be
explained by the interaction between the ambient material in the
molecular cloud and an outflow launched by the core.  Finally, the
3\,mm continuum emission detected towards Per-Bolo 45 is also
suggestive of a possible embedded source, given that 3\,mm continuum
emission was detected with CARMA towards Per-Bolo 58 and L1541-mm, two
supposedly ``starless'' cores that were subsequently found to be
protostellar \citep{Enoch10, Dunham11, Pineda11}.  None of the other
starless cores surveyed with CARMA by \citet{Schnee10} exhibited 3\,mm
continuum emission.

Given that Per-Bolo 45 is in the NGC 1333 cluster, it is important to
check that the outflow does not originate from another nearby
protostar.  The elongation of the SiO emission points back towards the
HH 7-11 group associated with SVS 13, and even further perhaps towards
IRAS 4, but each of these cores have well-known outflows going in
directions other than towards Per-Bolo 45.  For example, a CO (3-2)
map of NGC 1333 shows no outflow directed towards Per-Bolo 45 from any
other object \citep{Curtis11}.  Instead, Per-Bolo 45 seems to be
adjacent to a compact red lobe emanating from IRAS 7 to the northeast
and a very extended red lobe emanating from IRAS 2 in the southwest.
These redshifted lobes have velocity ranges of 12\,\kms\ to
18\,\kms\ and their blueshifted lobes range from -5\,kms\ to 3\,\kms,
both outside the range of velocities associated with the redshifted
and blueshifted \hcop\ emission from Per-Bolo 45.  It would also be
difficult to explain how an outflow from elsewhere in the NGC 1333
cluster could interact with the dense material in Per-Bolo 45 traced
by \nthp\ and \nhtd\ in such a way as to increase turbulence
immediately to the south-east of the 3\,mm continuum peak without also
affecting the area with nearly thermal line widths around it.  We
conclude that the molecular line observations presented in this paper
are best explained by the presence of a protostar embedded within
Per-Bolo 45.

\subsection{Embedded Source} \label{EMBEDDED}

A relationship between the internal luminosity of a young stellar
object (YSO) and its 70\,\micron\ flux is given by Equation 2 in
\citet{Dunham08}.  Per-Bolo 45 is not detected at 70\,\micron, but we
can use a 3\,$\sigma$ upper limit to its 70\,\micron\ flux to estimate
an upper limit to the internal luminosity of the embedded source.  We
find that the internal luminosity of Per-Bolo 45 is less than
$10^{-2}$\,\solum, similar to the upper limit for L1451-mm by
\citet{Pineda11} and lower by a factor of about 10 than the
luminosities of the embedded protostars observed by {\it Spitzer} with
published models shown in Table 1 of \citet{Dunham08}.  The upper
limit to the luminosity of the embedded source in Per-Bolo 45 is
consistent with the sensitivity limit of the c2d survey.  Assuming
that Per-Bolo 45 is protostellar, the lack of observed
70\,\micron\ flux would make this source a VeLLO by the definition
given in \citet{DiFrancesco07}.  A protostellar source with a disk
viewed edge-on might remain invisible at the observed sensitivity
limits, so it is possible that the internal luminosity of Per-Bolo 45
is higher than that of a VeLLO and it appears dim because of the
viewing angle.  Given its low 70\,\micron\ luminosity, the source
embedded in Per-Bolo 45 is also a plausible first hydrostatic core
candidate.

\section{Discussion} \label{DISCUSSION}

In this paper we identify Per-Bolo 45 as a core with an embedded
source, changing its classification from starless core \citep{Enoch08}
to VeLLO.  Other cores in the Perseus molecular cloud, previously
identified as starless due to their lack of near- and mid-infrared
emission \citep[e.g.][]{Enoch08, Sadavoy10} but subsequently found to
show evidence for protostellar activity, such as launching molecular
outflows, include Per-Bolo 58 \citep{Enoch10,Dunham11} , L1448 IRS2E
\citep{Chen10a}, and L1451-mm \citep{Pineda11}.

Out of the 11 ``starless'' cores in Perseus surveyed by
\citet{Schnee10} with interferometric observations of their 3\,mm
continuum, the only two cores detected (Per-Bolo 45 and Per-Bolo 58)
have since been shown to harbour embedded objects.  This fraction of
misidentified starless cores in Perseus, 2/11 or 18\%, is probably an
upper limit given that the cores in the \citet{Schnee10} sample were
chosen to have high surface brightnesses in the 1.1\,mm continuum map
of \citet{Enoch06}, and the presence of a protostar would increase the
peak flux of a dense core.

A lower limit to the number of cores in Perseus misidentified as
starless can be derived from the number of previously identified
starless cores and the number of newly identified protostars and
VeLLOs.  This estimate provides a lower limit because not every core
identified in wide-field dust continuum maps of Perseus has been
followed up with (sub)millimeter interferometric molecular line
observations to find outflows.  \citet{Sadavoy10} find 97 starless
cores and 46 protostellar cores in Perseus, using SCUBA
850\,\micron\ maps to identify the cores and {\it Spitzer} maps
between 3.6 and 70\,\micron\ to determine the starless or protostellar
status of each core.  \citet{Jorgensen07} and \citet{Enoch08} find
similar numbers of starless cores and protostars using similar
techniques.  Although the census of protostellar cores and starless
cores detected by {\it Herschel} has not yet been released for the
Perseus molecular cloud, a 2:1 ratio of prestellar cores to
protostellar cores in the Aquila rift was reported by
\citet{Konyves10} and \citet{Bontemps10}, in good agreement with the
pre-{\it Herschel} studies of Perseus described above.  Considering
the four newly-identified VeLLOs in Perseus, a lower limit of 4/97
``starless'' cores, or 4\%, have been misclassified.  Our upper limit
of 18\% would imply that 17 cores have been misclassified as starless.
We note that it is likely that not all of the dense cores in Perseus
have been identified, given that the published (sub)millimeter surveys
used to identify such cores have had limited sensitivity to faint and
extended structures.

The lifetime of starless cores can be estimated from the relative
number of starless cores and protostellar cores, given a reasonable
estimate of the protostellar lifetime and assuming that the rate of
star formation is constant over time.  \citet{Enoch08} found that a
mean lifetime for starless cores in nearby molecular clouds is $0.5
\pm 0.3$\,Myr, so the $\le$20\% change in the number of starless cores
in Perseus resulting from the discovery of embedded sources in
previously identified ``starless'' cores is roughly comparable to the
uncertainty coming from other systematics in the estimate of starless
core lifetimes.

Similarly, increasing the number of protostars increases the resultant
protostellar lifetime.  \citet{Evans09} found 87 Class 0/I protostars
in the Perseus molecular cloud, so adding 4-17 more sources into the
Class 0/I category would increase the Class 0/I lifetime by
$\sim$5-20\%.  This increase in the Class 0/I lifetime would not
significantly affect modelling of the expected accretion luminosity
averaged over the lifetime of a protostellar core, given the
significant uncertainties on both the observational and theoretical
sides of the problem, as summarized in \citet{Offner11}.

Many studies of nearby molecular clouds \citep[e.g., ][]{Motte98,
  Nutter07, Ward-Thompson07, Enoch08, Konyves10} have shown that the
starless core mass function (CMF) has a similar shape to the stellar
initial mass function (IMF).  These studies depend on the ability to
distinguish starless dense cores from protostellar dense
cores. \citet{Hatchell08}, \citet{Enoch08} and \citet{DiFrancesco10}
each found that their most massive cores tended to be protostellar,
which suggests that either prestellar cores have an upper mass limit,
consistent with gravitational stability inequalities (e.g., Jeans
analysis), or that the more massive prestellar cores have very short
lifetimes.  Furthermore, \citet{Sadavoy10b} found that the most
massive ``starless'' cores in Orion and Perseus had ambiguous infrared
emission towards them and thus, changed their classification to
``undetermined.''  If the most massive ``starless'' cores tend to be
misclassified VeLLOs, then the CMF will lack the massive core tail
found in the IMF.  This difference will steepen the starless CMF slope
with respect to the IMF slope, i.e., the two mass functions will have
different shapes.

Three of the VeLLOs in Perseus (Per-Bolo 45, Per-Bolo 58, and L1448
IRS2e) were included in the \citet{Sadavoy10b} investigation into the
gravitational stability of starless cores in Perseus.  None of these
sources were found to be particularly unstable.  Instead, they reside,
along with the majority of starless cores, in a mass range of 1-3
Jeans masses \citep[see][for details on how the stability analysis was
  performed]{Sadavoy10b}.  The only possible hint that these cores are
unique is that all three have relatively small sizes (compared with
other cores at similar level of Jeans instability).  That these cores
are not clearly distinguishable by a Jeans analysis further suggests
that deep mid-infrared and interferometric observations will be
required to uncover the majority of VeLLOs.  Given that only a few
``starless'' cores have been proven to be misclassified, and that
these cores do not stand out as being especially massive, the shape of
the CMF, as discussed above, has not been biased by the presence of
low-luminosity protostellar objects.

In a 3\,mm continuum survey of 11 ``starless'' cores, \citet{Schnee10}
found that only two (Per-Bolo 45 and Per-Bolo 58) were detected and
concluded that the density distribution in starless cores must be
smooth and not strongly peaked.  The subsequent discovery that both
cores are actually VeLLOs \citep[][this paper]{Enoch10,Dunham11}
strengthens the claim that starless cores in Perseus have smooth and
shallow density profiles.  A shallow inner density profile in starless
cores has been previously reported by several groups,
\citep[e.g.,][]{Ward-Thompson94, Ward-Thompson99, Shirley00}.  Had the
starless cores in our Perseus sample been in the process of
fragmentation, we would have been able to detect the resultant
fragments in the dust emission maps.

Although {\it Spitzer} surveys were more sensitive than those
conducted with previous instruments and found embedded objects in
cores that had been identified as starless \citep{Young04,
  Kauffmann05, Bourke06}, it is also true that deeply embedded
protostars can be missed at wavelengths less than
12\,\micron\ \citep{Jorgensen08} by surveys like c2d \citep{Evans03}.
In addition, \citet{Dunham08} have found that there is likely to be a
population of VeLLOs too faint to have been detected in current
surveys.  New {\it Herschel} observations will be able to find fainter
protostars than {\it Spitzer} was able to detect, and in the Aquila
region seven of the $\sim$50 Class 0 protostars detected by {\it
  Herschel} were missed by {\it Spitzer} \citep{Bontemps10}.  Still,
\citet{Bontemps10} reported that compact sources in Aquila were
identified down to the {\it Herschel} 70\,\micron\ detection limit,
implying that a population of even fainter sources have yet to be
discovered.  Given that low-luminosity embedded objects can be found
through high-resolution observations of their dust continuum emission
and molecular outflows, we suggest that interferometric
(sub)millimeter observations are a promising method for determining
what fraction of ``starless'' cores have been misclassified.

\section{Summary} \label{SUMMARY}

In this paper, we report on CARMA maps of the 3\,mm continuum and
3\,mm window spectral lines of \nhtd, SiO, \hcop, HNC, and \nthp.  Our
main results are:
\begin{itemize}
\item{Despite non-detections in {\it Spitzer} maps at 3.6-24\,\micron,
  Per-Bolo 45 is a protostellar core, as inferred from large line
  widths, Doppler-shifted emission, and the presence of SiO emission.}
\item{There are at least four cores in Perseus, previously identified
  as starless, that recent observations have shown to contain
  low-luminosity embedded objects with molecular outflows
  \citep{Chen10a, Enoch10, Dunham11,Pineda11}.  If this result can be
  generalized to other nearby molecular clouds, we estimate that the
  lifetime of starless cores has been overestimated by 4-18\% and the
  lifetime of Class 0/I protostars has been underestimated by 5-20\%.
  These changes are within the previously published uncertainties of
  starless and protostellar core lifetimes \citep{Enoch08, Evans09}.}
\item{Although recent infrared surveys of nearby molecular clouds have
  made great progress towards classifying starless and protostellar
  cores \citep[e.g.,][]{Hatchell07a, Jorgensen07, Jorgensen08,
    Enoch08, Dunham08, Evans09, Sadavoy10, Konyves10, Bontemps10}, we
  suggest that an improved census will require high-resolution
  (sub)millimeter observations to survey known cores and identify
  embedded low-luminosity objects and outflows.}
\end{itemize}
\acknowledgments

We thank our anonymous referee for comments that have significantly
strengthed this paper.  JDF acknowledges support by the National
Research Council of Canada, the Canadian Space Agency (via a SSEP
Grant), and the Natural Sciences and Engineering Council of Canada
(via a Discovery Grant).  Support was provided to ME by NASA through
the {\it Spitzer Space Telescope} Fellowship Program.  DJ is supported
by a Natural Sciences and Engineering Research Council of Canada
(NSERC) Discovery Grant.  We thank the CARMA staff, students and
postdocs for their help in making these observations.  Support for
CARMA construction was derived from the Gordon and Betty Moore
Foundation, the Kenneth T. and Eileen L. Norris Foundation, the
Associates of the California Institute of Technology, the states of
California, Illinois and Maryland, and the National Science
Foundation.  Ongoing CARMA development and operations are supported by
the National Science Foundation under a cooperative agreement, and by
the CARMA partner universities.

{\it Facilities}: CARMA

{}

\begin{deluxetable}{lccc} 
\tablewidth{0pt}
\tabletypesize{\scriptsize}
\tablecaption{CARMA Observations \label{MMOBSTAB}}
\tablehead{
 \colhead{Line}       & 
 \colhead{$\nu$\tablenotemark{1}} & 
 \colhead{beam size}  & 
 \colhead{rms\tablenotemark{2}} \\
 \colhead{}           &
 \colhead{GHz}        &
 \colhead{\arcsec}    &
 \colhead{K}}
\startdata
NH$_2$D (1$_{1,1}$ - 1$_{0,1}$ F=2-2) & 85.9262703\tablenotemark{3} & 7.4$\times$5.5 & 0.45 \\
SiO (2-1)                         & 86.84696\tablenotemark{4}   & 7.6$\times$5.7 & 0.42 \\
HCO$^+$ (1-0)                     & 89.188523\tablenotemark{4}   & 6.9$\times$5.3 & 0.42\\
HNC (1-0)                         & 90.663568\tablenotemark{4}  & 7.2$\times$5.4 & 0.42 \\
N$_2$H$^+$ (1-0) F$_1$=2-1 F=3-2   & 93.173777\tablenotemark{5}  & 7.0$\times$5.3 & 0.46 \\
C$^{34}$S (2-1)                    & 96.4129495\tablenotemark{4} & 6.9$\times$4.9 & 0.47
\enddata
\tablenotetext{1}{Rest frequency}
\tablenotetext{2}{Per beam, in a 0.1\,\kms\ channel}
\tablenotetext{3}{\citet{Tine00}}
\tablenotetext{4}{\citet{Schoier05}}
\tablenotetext{5}{\citet{Lovas92}}
\end{deluxetable}

\end{document}